\documentclass[%
reprint,
superscriptaddress,
amsmath,amssymb,
aps,
prl,
floatfix
]{revtex4-2}

\usepackage{multirow}
\usepackage{graphicx}
\usepackage{dcolumn}
\usepackage{bm}
\usepackage[version=3]{mhchem}
\usepackage{nicefrac}

\usepackage{color}
\usepackage{dcolumn}
\usepackage{amssymb}
\usepackage{url}
\usepackage{hyperref}
\usepackage{xfrac}
\usepackage{tabularx}
\usepackage{xspace}
\usepackage{amsmath}
\usepackage[normalem]{ulem}
\usepackage{mathtools}
\usepackage{physics}
\usepackage{dsfont}

\def\be{\begin{equation}}
\def\ee{\end{equation}}
\def\bea{\begin{eqnarray}}
\def\eea{\end{eqnarray}}
\def\ba{\begin{array}}
\def\ea{\end{array}}

\def\S{\hat{\mathbf{S}}}

\begin{document}
\title{Where is the spin liquid in maple-leaf quantum magnet?}
\author{Pratyay Ghosh}
\email{pratyay.ghosh@epfl.ch}
\affiliation{Institute of Physics, Ecole Polytechnique Fédérale de Lausanne (EPFL), CH-1015 Lausanne, Switzerland}


\begin{abstract}
We investigate the possibility of exotic phenomena, viz. quantum spin liquid (QSL) or deconfined quantum critical point (DQCP), in the spin-$\frac{1}{2}$ Heisenberg model on the maple-leaf lattice, a geometrically frustrated system formed by hexagons (coupling $J_h$), triangles (coupling $J_t$), and dimers (coupling $J_d$). We identify one promising region, given by $J_h > 0$ and $J_t, J_d < 0$, for hosting enticing physics. In this region, the quantum phase diagram of the system exhibits an interplay between N\'eel order and a gapped dimerized singlet phase. This arrangement holds the possibility of harboring a QSL and a DQCP. Using bond-operator mean-field theory and density matrix renormalization group calculations, we delve into this uncharted territory, revealing tantalizing evidence of the existence of a QSL phase and highlighting its potential as a platform for DQCP.
\end{abstract}

\maketitle

Modern condensed matter physics extensively explores the low-temperature behavior of matter, focusing on the competition between qualitatively distinct ground states and the associated phase transitions~\cite{Sachdev1999,Diepbook,frustrationbook}. This line of inquiry originates from the seminal contributions of Landau, who incepted a paradigm centered around the symmetries of many-body systems~\cite{TERHAAR1965}. While a microscopic Hamiltonian inherently possesses specific symmetries, the ground state (GS) is not required to adhere to all of them; some symmetries may be spontaneously broken. A symmetry-broken state markedly differs from one that preserves the symmetries. Landau proposed the notion of an order parameter to quantify the extent of such broken symmetry and characterize the symmetry-broken phase. Within the same parameter space, the system may switch between symmetry-broken and symmetry-preserving phases, depending on the specific values of the system's parameters. This changeover must then involve a phase transition, the properties of which are linked to the fluctuations in the order parameter~\cite{landau2013statistical,Wilson1974}.

This concept, developed by Landau and Ginzburg, effectively describes a broad spectrum of phase transitions, yet a notable fraction of transitions is beyond this framework. A particularly intriguing instance is the second-order phase transition between two distinct symmetry-broken Landau-ordered phases, exemplified in the $J_1$-$J_2$ model on the square lattice, where a continuous phase transition was observed between an antiferromagnetic (AFM) order (breaking spin-rotation symmetry) and a valence-bond solid (VBS) state (breaking lattice-rotation symmetry)~\cite{Zhitomirsky1996,BOT-Sachdev,Capriotti2000,Mambrini2006,Doretto2014}. To explain such transitions, the concept of a deconfined quantum critical point (DQCP) was proposed~\cite{deconfined,Senthil2004,Senthil2023}. Notably, in some cases, an intermediate quantum spin liquid (QSL) phase concocts between the magnetic order and the VBS~\cite{Jiang2012,Gong2013,Gong2014,Lee2019,Yang2022,Wang2018a,Ferrari2020}, sometimes originating from a nearby DQCP~\cite{Liu2022,Yang2022}. 

\begin{figure}[t]
\includegraphics[width=0.7\columnwidth]{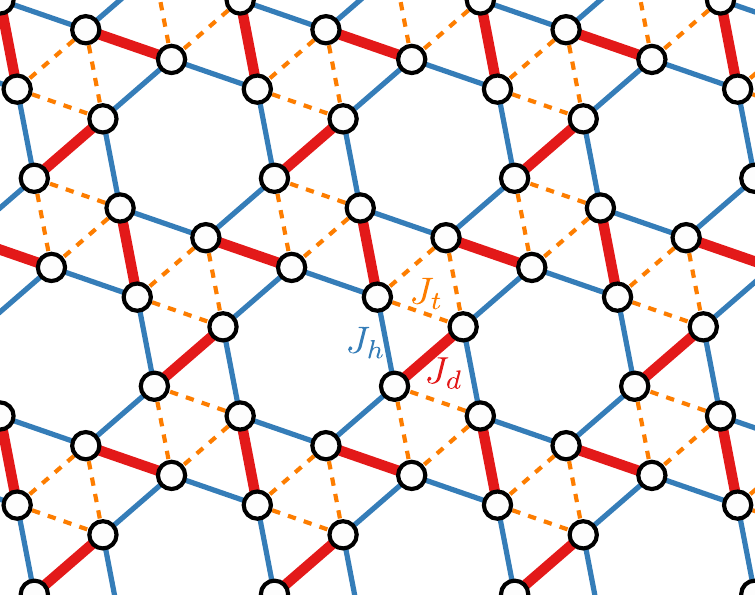}
\caption{The Maple-leaf model (MLM) given in~\eqref{eq-hamil} with three symmetry-inequivalent bonds indicated by thick red ($J_d$), blue ($J_h$), and dotted orange ($J_t$). The model admits an exact dimer singlet GS when $J_d$ is sufficiently larger than $J_t=J_h>0$~\cite{Ghosh2022,Ghosh2023}.} \label{fig-lattice}
\end{figure}

A prominent scenario recently explored in the context of a deconfined criticality and QSL is the famous Shastry-Sutherland model (SSM)~\cite{Shastry1981}. Within the SSM, a QSL phase is found in a narrow region between the plaquette VBS and the N\'eel ordered phases~\cite{Lee2019,Yang2022,Viteritti2023}. This QSL phase is proposed to emanate from a nearby DQCP. Adding to this discourse is the maple-leaf model (MLM), a nearest-neighbor Heisenberg model on the maple-leaf lattice (MLL)~\cite{Betts1995} illustrated in Fig.~\ref{fig-lattice}, which recently regained attention following the findings in Ref.~\cite{Ghosh2022}, highlighting its significance alongside the (SSM) as the only other 2D lattice of uniform tiling that allows for an exact dimer singlet GS. The MLM Hamiltonian is defined as follows.
\be\label{eq-hamil}
\hat{H}=J_h\sum_{\langle ij\rangle_h}\S_{i}\cdot\S_{j}+J_t\sum_{\langle ij\rangle_t}\S_{i}\cdot\S_{j}+J_d\sum_{\langle ij\rangle_d}\S_{i}\cdot\S_{j}.
\ee
Here, the sums $\langle\rangle_k$ run over nearest-neighbors connected by a bond-type $k$ with a coupling strength $J_k$ (Fig.~\ref{fig-lattice}). There are three symmetry-inequivalent bonds: with coupling strength $J_h$ on the solid hexagons, $J_t$ on the dashed triangles, and $J_d$ on the thick dimers. $\S_i$ denotes the spin-$1/2$ operator on site $i$. For AFM $J_h=J_t\leq 2J_d$, the model is proven to host a product dimer singlet GS~\cite{Ghosh2022,Ghosh2023}. The model has been a subject of several numerical investigations~\cite{Farnell2011,Schmalfuss2002,Jahromi2020,Makuta2021,Gresista2023,Beck2024}, most of which found a canted $120^\circ$ magnetic order for $J_d=0$, $J_h=J_t>0$. Despite these similarities with the SSM, namely an exact singlet phase and a magnetic order which naturally raises expectations for a VBS between the exact singlet and magnetic order, facilitating a QSL or a DQCP in the proximity, no compelling evidence was found for it~\cite{Farnell2011,Gresista2023,Beck2024}. In this letter, we argue that such an intermediate phase is unlikely. Our arguments extend further to identify regions in the MLM that may manifest these exotic physics, supported by both semi-analytic and numerical calculations. Moreover, we connect the parameter space to potential material realizations.

We first address the prospect of a QSL in MLM when $J_h=J_t\neq J_d>0$. Previous works~\cite{Farnell2011,Ghosh2022} have found, in the classical limit ($\vec{S}\to\infty$), a unique canted $120^\circ$ magnetic order for $J_d\in[0,\infty)$. In the quantum case, the system maintains this magnetic order for a substantial $J_d$~\cite{Farnell2011,Farnell2014,Schmalfuss2002,Gresista2023,Beck2024}. For such a stable, non-degenerate state, the quantum fluctuations is reluctant to engender a second-order phase transition out of the magnetically ordered phase at a finite $J_d$; instead, such a transition must be first-order. Likewise, the phase transition, from large $J_d$ side, out of the exact singlet phase, typically, occurs at first-order~\cite{Koga2000,Corboz2013,Lee2019,Ghosh2023a,Ghosh2023c}. Consequently, one suspects a direct first-order transition between the $120^\circ$ magnetic order and the exact singlet phase, as also suggested by Refs~\cite{Farnell2011,Beck2024}. This situation is different in SSM, where the classical spin system undergoes a second-order phase transition from N\'eel to a spiral phase as dimer interactions strengthen~\cite{Shastry1981} allowing for a continuous transition out of the N\'eel phase in the quantum case.

\begin{figure}
\includegraphics[width=0.95\columnwidth]{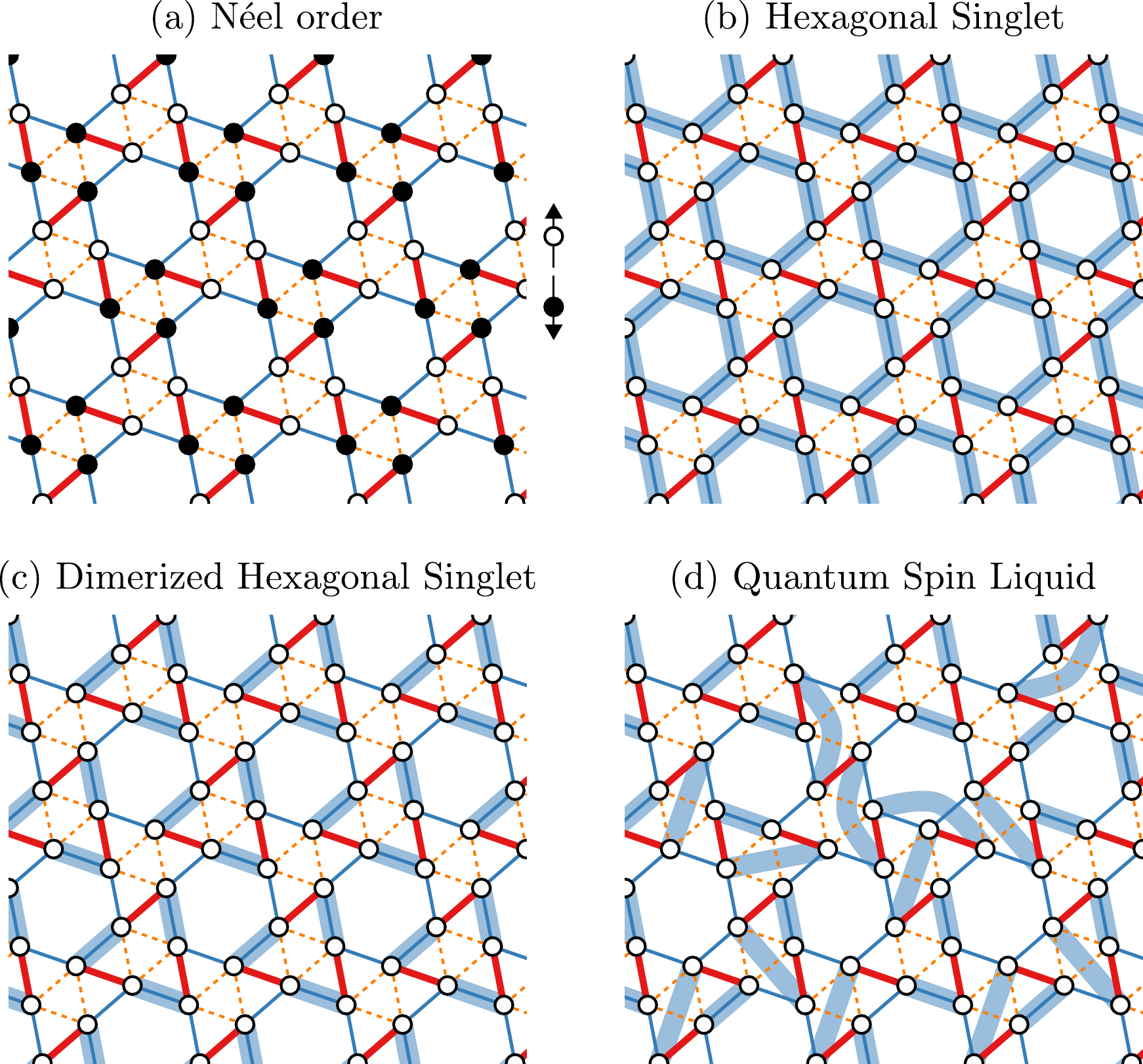}
\caption{Possible GSs of the MLM: (a) N\'eel order, (b) hexagonal singlet with strong and uniform singlet amplitudes (shown in thick light-blue) on $J_h$ hexagons, (c)
dimerized hexagonal singlet with dimerization on the $J_h$ hexagons, and (d) quantum spin-liquid phase. The MLL is classified under the $p6$ wallpaper group, featuring a $C_3$ symmetry with respect to the center of the $J_t$ triangles and a $C_6$ symmetry around the $J_h$ hexagons. Thus, the dimerized hexagonal singlet breaks lattice-rotation symmetry, and the hexagonal singlet and the spin-liquid (depicted only by short-range singlets) do not break any symmetry.} \label{fig-states}
\end{figure}

We, now, discuss two possible VBS states that we might encounter in the MLM's parameter space. For $J_t=0$ and $J_h=J_d>0$, the model is identical to an AFM Heisenberg model on a honeycomb lattice, with the N\'eel ordered GS depicted in Fig.~\ref{fig-states} (a). Introducing $J_t>0$ includes a subset of second-neighbor interactions. The $J_1$-$J_2$ model on the honeycomb lattice has hinted at the possibility of a QSL and a DQCP emerging between the N\'eel state and the lattice symmetry-breaking plaquette VBS, appearing for $J_2\gtrsim 0.25J_1$~\cite{Oitmaa2011,Reuther2011,Ganesh2013,Gong2013}. However, the scenario differs in the MLM. In there, the plaquette VBS state resembles the one illustrated in Fig.~\ref{fig-states} (b), forming strong singlets on the $J_h$ hexagons~\cite{Beck2024}. We call this the \emph{hexagonal singlet} state. As $J_d$ and $J_h$ are not symmetry-related, this plaquette state doesn't break any symmetry. Thus, even if a continuous phase transition materializes between the N\'eel state and the hexagonal singlet, it lies perfectly within the Landau-Ginzburg paradigm, offering no guarantee of an intermediate QSL. By the same reasoning, a second-order transition out of the canted $120^\circ$ phase and into a VBS with strong singlets on the $J_d$ bonds is possible but might lack interesting physics.

A VBS state that meets the criteria for exotic criticality is depicted in Fig.~\ref{fig-states} (c), where the strong singlet amplitudes are highlighted by light-blue ellipses, and we term this state the \emph{dimerized hexagonal singlet}~\cite{Beck2024}. This state breaks lattice rotation symmetry through dimerization. A second-order phase transition between this VBS and a magnetic order is Landau-forbidden and should occur at a DQCP, and in its vicinity, we can expect a QSL originating from the DQCP. However, the question remains: where does this VBS phase manifest? The answer lies in $J_h>0$, $J_t=0$, and $J_d\to -\infty$ limit of the MLM. Here, the spin-$1/2$s astride a strong FM dimer bond project onto the spin-$1$ sector, transforming the system into an effective AFM spin-$1$ kagom\'e, the GS of which is a trimerized singlet~\cite{Liu2015,Changlani-Spin-1_Kagome,Ghosh-Spin-1_Kagome}. This state breaks the symmetry between the up and down triangles of kagom\'e, forming stronger singlets on one of them. In our model where the spin-$1$s are split into two spin-$1/2$ across a strong ferromagnetic (FM) bond, the trimerized state of spin-$1$ kagom\'e translates to a dimerized state on the $J_h$ hexagons~\cite{Ghosh_Hida_Model_of_Kagome}. This construction is analogous to the Affleck-Kennedy-Lieb-Tasaki (AKLT) state for the AFM spin-$1$ chain~\cite{AKLT,Haldane1983}. In both instances, one interprets the spin-$1$ as a composite of two spin-$1/2$s, the spin-$1/2$s form a VBS state, and then two spin-$1/2$ projects out a spin-$1$.

In the first-order perturbation from strong FM $J_d$ limit, the effective interactions between these spin-$1$-like entities are $\sim J_h + J_t$. Therefore, depending on the values of $J_h$ and $J_t$, the system will behave as either a FM or an AFM spin-$1$ kagom\'e. Since we seek dimerization at large $|J_d|$, we must ensure $J_h + J_t>0$. To drive the system towards N\'eel order [Fig.~\ref{fig-states} (a)] for $|J_d|\approx 0$, we set $J_h > 0$ and $J_t < 0$. With $J_t$ and $J_h$ satisfying the conditions, varying $J_d$ from $0$ to a significantly large negative value, we anticipate observing phase transition(s) from the N\'eel phase to the dimerized hexagonal singlet phase. We will henceforth refer to this case as the special maple-leaf model (sMLM). In the sMLM, the sole source of frustration is the FM $J_d$ bonds. We can draw an analogy with the SSM;  both models are frustrated only by dimer interactions, and like the SSM, the sMLM with classical spins demonstrates a second-order phase transition from the N\'eel phase to a spiral phase (details in the Supplemental Material (SM)~\cite{supp} which also contains the additional references~\cite{Luttinger1946,Lapa2012}). 

Note that the spin-$1/2$ sMLM devises an intriguing property when $|J_t|=J_h$, wherein it can accommodate exact eigenstates; the entire multiplet can be obtained by repeated application of $\sum_i\hat{S}_i^-$ on the fully polarized state. This phenomenon is exclusive to the MLL and does not crop up in the SSM unless the symmetry of the inter-dimer bonds is explicitly broken. In the case of the MLL, these eigenstates naturally arise due to the lack of symmetry relation between $J_t$ and $J_h$. These eigenstates persist within the middle spectrum for finite $|J_d|$. We presume that these states may relate to the exact excited states of the AKLT model~\cite{Scar-AKLT}, with the potential for many-body localization and quantum scars~\cite{scarMaple}.

\begin{figure}[t]
\includegraphics[width=0.9\columnwidth]{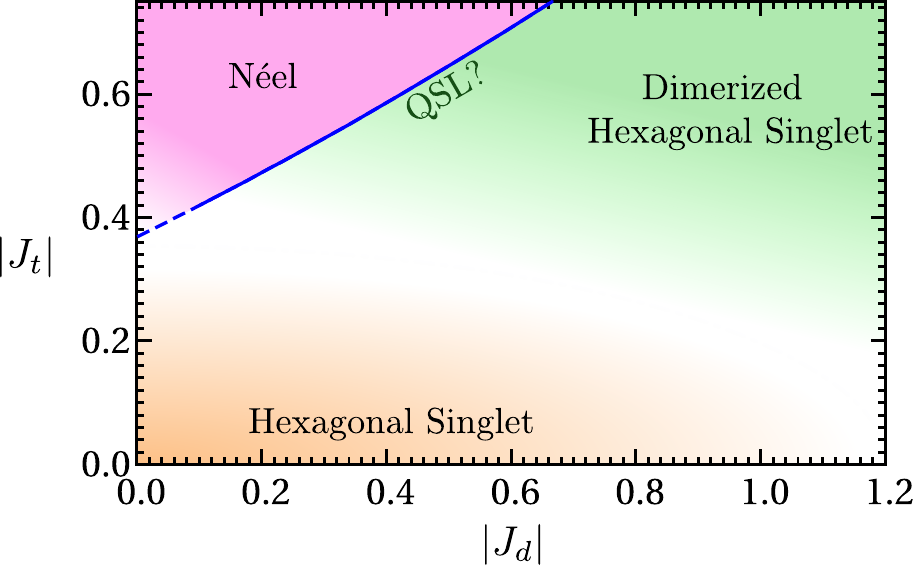}
\caption{The quantum phase diagram of the sMLM derived from bond-operator mean-field analysis. The gapped dimerized singlet phase is stable for a range of $J_d$. Upon lowering of $|J_d|$, the system undergoes a continuous transition to the gapless N\'eel ordered phase. The hexagonal singlet phase should appear for small $J_t$ and $J_d$.} \label{fig-tmft}
\end{figure}

Now, we address the final question: Does the MLM exhibit an exotic phase? The anticipated GS for the sMLM with a dominant FM $J_d$ is dimerized. For such a state, the bond-operator mean-field theory~\cite{BOT-Sachdev} can provide an effective low-energy theory, represented by one singlet and three triplet bosons residing on a bond. The bond-operator representation can describe systems that respect the spin rotational symmetry as well as magnetic order. This approach has yielded crucial insights into the low-energy physics of various magnetic systems~\cite{Kotov1998,Matsumoto2002,Kumar2010,Normand2011,Ghosh_Hida_Model_of_Kagome,Ghosh-Spin-1_Kagome,Nawa2019}. In fact, in Ref.~\cite{BOT-Sachdev}, Sachdev and Bhatt studied the AFM $J_1$-$J_2$-$J_3$ Heisenberg model on the square lattice, now recognized to host a QSL phase~\cite{Jiang2012,Wang2018a,Ferrari2020} emerging from a DQCP~\cite{Liu2022} and sandwiched between the N\'eel and valence-bond solid (VBS) phases. The bond-operator mean-field theory can not find a QSL, rather it just indicates a continuous phase transition between the two Landau orders. Despite that, this approach provides an initial indication of the exotic criticality that the system can showcase. 

\begin{figure}[b]
\includegraphics[width=0.85\columnwidth]{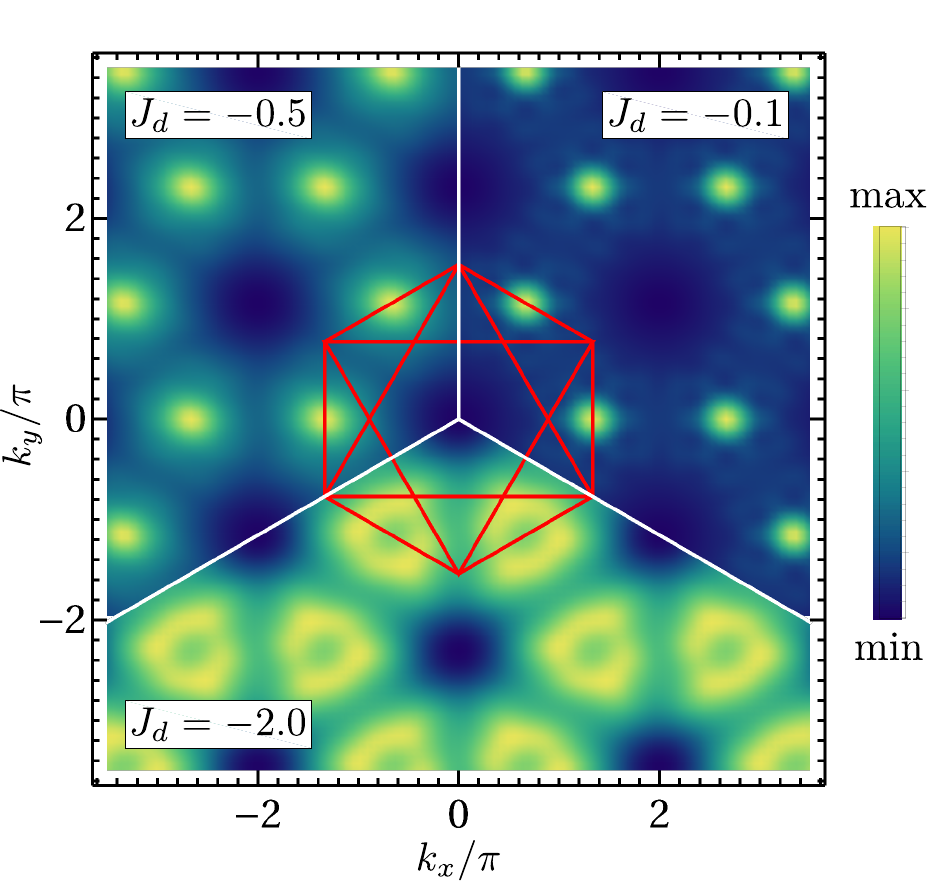}
\caption{The static spin structure factors derived from DMRG for N\'eel order at $J_d=-0.1$, the intermediate phase, possibly a QSL, for $J_d=-0.5$, and dimerized singlet phase at $J_d=-2.0$. We have fixed $J_h=1$ and $J_t=-0.5$. The first three Brillouin zones (cluster geometry in \cite{supp}) are indicated.} \label{fig-Sq}
\end{figure}

We employ the bond operator representation on the alternating $J_h$ bonds, indicated by the thick-blue ellipses in Fig.~\ref{fig-states} (c). Our approach assumes a singlet background (a product state of singlets on the specified $J_h$ bonds) as a mean-field, incorporates the effect of triplon (dispersing triplet) excitations while neglecting any triplon-triplon interaction (details in the SM~\cite{supp}). The mean-field analysis unveils the expected gapped dimerized GS for large $|J_d|$. This dimerized phase remains stable for intermediate values of $|J_d|$ when $|J_t|\gtrsim 0.4$ ($J_h=1$ throughout the remainder of this letter). Eventually, for sufficiently small $J_d$, it undergoes a continuous phase transition to the gapless N\'eel order depicted in Fig.~\ref{fig-states} (a). Such a phase transition is beyond the Landau-Ginzburg paradigm; and warrants for a DQCP and a QSL phase to appear between these two phases, similar to the AFM $J_1$-$J_2$-$J_3$ model on the square lattice. The quantum phase diagram in Fig.~\ref{fig-tmft} includes a hexagonal singlet phase, expected for small $J_t$ and $J_d$. While we do not explicitly study the phase boundary between the hexagonal singlet and the other phases here, it can also be an intriguing aspect to explore.

Using density matrix renormalization group (DMRG) methods~\cite{White1992} utilizing the iTENSOR library~\cite{itensor}, we also find indications of a possible intermediate phase between the N\'eel and the dimerized phase (details in SM~\cite{supp}). The indication of the intermediate phase is evident in the static structure factor, $S(\mathbf{k})\sim\sum_{ij}\langle\S_i\cdot\S_j\rangle\exp(i \mathbf{k}\cdot\mathbf{r}_{ij})$ (where $\mathbf{r}_{ij}$ is the vector connecting sites $i$ and $j$), shown in Fig.~\ref{fig-Sq} where we show the results for $(J_h,J_t)=(1,-0.5)$. The N\'eel phase is stable for $|J_d|\lesssim 0.35$~\cite{supp} and produces Bragg peak-like characters, whereas, for $J_d=-0.5$, no sharp peaks are observed which are notably different from the features corresponding to the dimerized phase at $J_d=-2$. Determining the nature of the GS in the intermediate region is not the scope of this letter. While a QSL phase might be expected, it is not a necessity. Any phase that does not break any symmetries, e.g. hexagonal singlet phase, could fit the description. 

\begin{figure}[t]
\includegraphics[width=0.85\columnwidth]{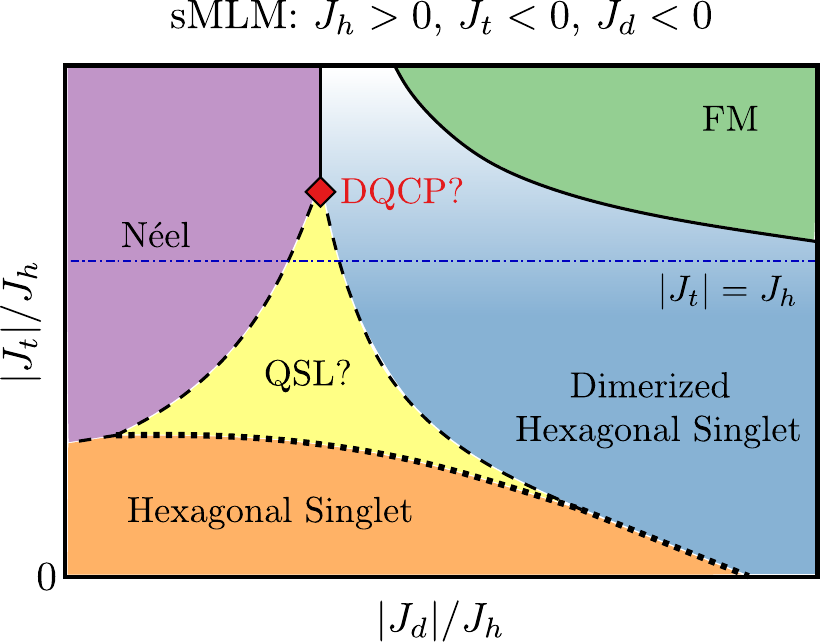}
\caption{A speculative phase diagram of the sMLM. The solid (dashed) lines represent first-order (second-order) phase transitions. We anticipate that the system hosts a QSL emerging out of a DQCP between the N\'eel and the dimerized hexagonal singlet state. As an alternative possibility, the QSL phase might also extend to $J_d\to\infty$. The $|J_t|=J_h$ of sMLM is a special case that accommodates exact singlet eigenstates. The nature of the phase transition out of the hexagonal singlet phase is mostly unknown.} \label{fig-QPD-full}
\end{figure}

As a concluding discussion, we present a speculative phase diagram of the sMLM in Fig.~\ref{fig-QPD-full}. The dimerized hexagonal singlet state should seemingly be stable below the $|J_t|=J_h$ line when $J_d$ is sufficiently strong. From the bond operator theory, where it is appropriately applicable, i.e. $|J_t|<J_h$, the phase transition out of the dimerized phase is second order in nature. However, for $|J_t|>J_h$, the effective system of spin-$1$'s becomes FM. We do not know the GS here, but we expect the dimerized singlet state should continuously connect to that. There are two possibilities, however, for the phase transition(s) between the N\'eel order and the magnetic disorder in this region: (i) the QSL prevails as $J_t\to-\infty$. (ii) the QSL ends at a DQCP, and from there, a first-order phase boundary between the putative phases extends to $J_t\to-\infty$. Confirming these speculations is numerically quite a challenge, which we keep as a future endeavor. Our aim here is to point out that within the MLM framework with a special combination of the interactions--termed the special maple-leaf model (sMLM)~\footnote{For strong AFM $J_t$, the MLM relates to the AFM star lattice, the GS of which is debated between a VBS and Resonating VBS state~\cite{Yang2010,Jahromi2018}. For $J_h=J_t>0$ and $J_d<0$, the classical canted $120^\circ$ state is expected to be stable and should evolve to the $\sqrt{3}\times\sqrt{3}$ magnetic order of spin-$1$ kagom\'e~\cite{Messio2011} as $J_d\to-\infty$.}--there is the plausibility of intriguing physical phenomena driven by the intricacies of DQCP and QSL which warrants further investigations.

There are several realizations of maple-leaf spin system, both natural and synthetic~\cite{Hawthorne1993,Olmi1995,Cave2006,Aliev2012,Kampf2013,Mills2014}, among which, only the magnetic properties of spangolite (\ce{Cu6Al(SO4)(OH)12Cl.3H2O})\cite{Fennell2011} and bluebellite (\ce{Cu6IO3(OH)10Cl})~\cite{Haraguchi2021,Ghosh2023b} have been investigated. Interestingly, both spangolite and bluebellite share a similarity with sMLM, as they both feature FM $J_d$ bonds~\cite{Fennell2011,Ghosh2023b}. In the case of bluebellite, the strength of these FM bonds imposes intra-layer effective spin-$1$ kagom\'e physics, composing a GS akin to our dimerized state~\cite{Ghosh2023b}. While the other interactions in bluebellite and spangolite differ from the sMLM, the vast parameter space capable of hosting QSL makes it plausible to discover a maple-leaf compound with a QSL. Further exploration of these compounds could uncover palpable novel quantum phenomena.

\textit{Acknowledgment:} The author thanks Fr\'ed\'eric Mila and Ronny Thomale for useful discussions.


%

\end{document}


\title{Supplementary Materials: Where is the spin liquid in maple-leaf quantum magnet?}
\author{Pratyay Ghosh}
\email{pratyay.ghosh@epfl.ch}
\affiliation{Institute of Physics, Ecole Polytechnique Fédérale de Lausanne (EPFL), CH-1015 Lausanne, Switzerland}


\maketitle

\section{LUTTINGER-TISZA ANALYSIS} 
\begin{figure}
\includegraphics[width=0.8\columnwidth]{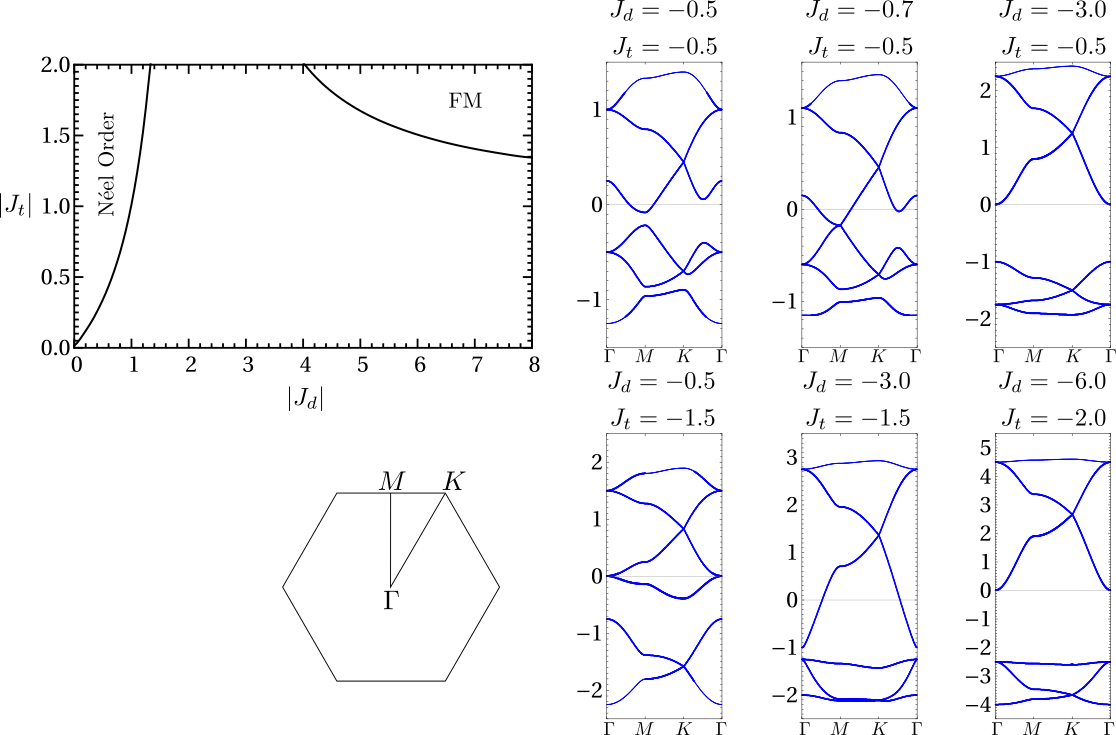}
\caption{The classical phase diagram of sMLM and the band structure in different regions of the parameter space obtained using Luttinger-Tisza analysis.} \label{fig-CPD}
\end{figure}

We ascertain the ground state of the sMLM through the application of the Luttinger-Tisza (LT) method~\cite{Luttinger1946}. The energy of a general classical spin system is expressed as follows:

\be
E=\frac{1}{2}\sum_{i,j}\sum_{k,l}J_{ij}\left(\R_{i}^{k}-\R_{j}^{k}\right)\mathbf{S}_i\left(\R^k\right)\mathbf{S}_j\left(\R^l\right),
\ee

Here, the unit cell positions are denoted by \(\R^k\), and the \(i\)-th basis site of the unit cell is labeled by \(\R_i^k\). The Fourier transformation yields:

\be
E=\frac{1}{2}\sum_{i,j}\sum_{k,l}J_{ij}\left(\k\right)\mathbf{S}_i\left(-\k\right)\mathbf{S}_j\left(\k\right),
\ee

with

\be
J_{ij}\left(\k\right)=\sum_{kl}J_{ij}\left(\R_{i}^{k}-\R_{j}^{k}\right)\exp[-i\k\cdot\left(\R_{i}^{k}-\R_{j}^{k}\right)].
\ee

First, the Fourier components of the interaction matrix, \(J_{ij}\left(\k\right)\), are diagonalized to obtain the lowest eigenvalue(s), \(\lambda_{min}(\k)\). Subsequently, one uses the fact that the energy per spin of any spin configuration satisfies the bound \(e\ge\lambda_{min}(\k)\).

A global ground state is identified if there exists a spin configuration that can be decomposed into a linear combination of only the 'optimal' LT eigenvectors corresponding to these eigenvalues~\cite{Lapa2012}. This occurs when the "strong constraint" of unit length for the spins,

\be
|\mathbf{S}_i|=1,
\ee

does not conflict with the optimal eigenvectors, which generally have entries with different amplitudes. In cases where non-optimal modes need to be admixed, typical for non-Bravais lattices, the LT approach provides a useful estimate for possible ground states.

The classical phase diagram for \(J_h=1\) and \(J_d,J_t<0\) obtained through the Luttinger-Tisza analysis is depicted in Fig.~\ref{fig-CPD}. The strong constraint is satisfied for both the N\'eel order and the FM phase. In the other phase, the minimum of the LT eigenvalues occurs at an incommensurate wave vector [see the eigenvalues of the interaction matrix along lines in the Brillouin zone in Fig.~\ref{fig-CPD}]. Additionally, note the appearance of FM and AFM kagome-like band structures for \(|J_t|>1\) and \(|J_t|<1\), respectively, with strong \(|J_d|\).

\section{Details of bond-operator mean-field theory}
\begin{figure}
    \centering
    \includegraphics[width=0.4\columnwidth]{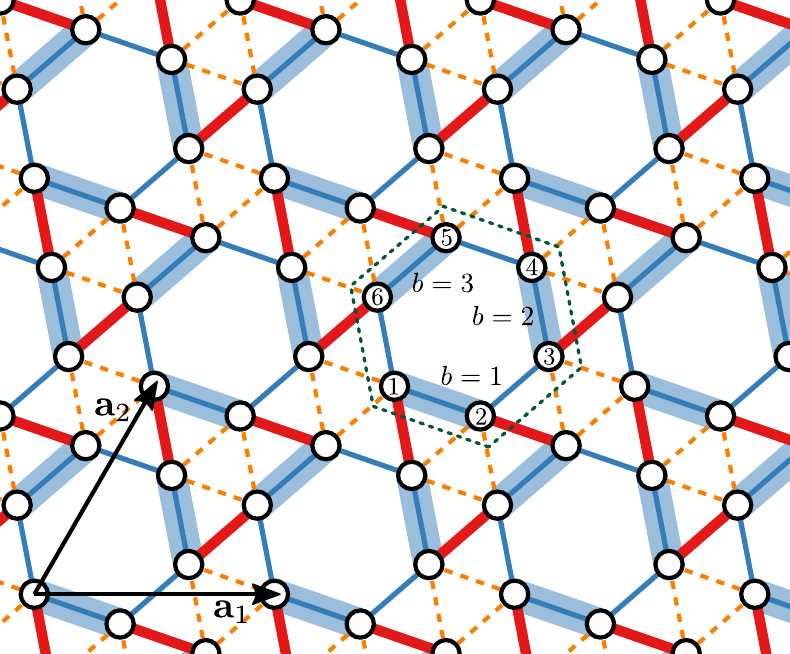}
    \caption{The dimerized hexagonal singlet ground state used in our bond-operator mean-field calculation. The singlets are forming on the alternate $J_h$ bonds. The indexing of the bonds and the sites in the unit cell are also marked.}
    \label{fig:S2}
\end{figure}
To comprehend the effective low-energy physics of the dimerized singlet ground state in the system, we begin by considering a $J_h$ hexagon as our unit cell. We focus exclusively on the three $J_h$ bonds, forming our elementary block, as illustrated by $\begin{gathered}\includegraphics[scale=0.14]{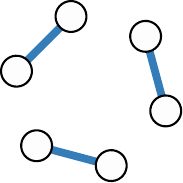}\end{gathered}$. The corresponding Hamiltonian is given by:
\bea\label{eq-hamil-tri}
\mathcal{H}_{\begin{gathered}\includegraphics[scale=0.14]{./t1.pdf}\end{gathered}}&=&J_h\sum_{b=1,2,3}\left(\mathbf{S}_{2b-1}\cdot\mathbf{S}_{2b}\right)\\
&=&J_h\left(\mathbf{S}_1\cdot\mathbf{S}_2+\mathbf{S}_3\cdot\mathbf{S}_4+\mathbf{S}_5\cdot\mathbf{S}_6\right),
\eea
where $b=1,2,3$ denotes the bond index (see Fig.\ref{fig:S2}). The ground state of this Hamiltonian is a product state of singlets forming on the $1\text{-}2$, $3\text{-}4$, and $5\text{-}6$ spin pairs. This characteristic allows us to utilize the bond-operator formalism~\cite{BOT-Sachdev} to express the spin operators as:
\begin{subequations}
\be
S_{2b-1}^\alpha=-\frac{1}{2}\left(\hats_{b}\hat{t}_{b}^{\alpha\dagger}+\hats_{b}^{\dagger}\hat{t}_{b}^\alpha\right)\\
\ee 
\be
S_{2b}^\alpha=\frac{1}{2}\left(\hats_{b}\hat{t}_{b}^{\alpha\dagger}+\hats_{b}^{\dagger}\hat{t}_{b}^\alpha\right).
\ee
\end{subequations}
When formulating the above representation, the basis includes the singlet $|s_b\rangle$ and three triplets $|t^{m}_b\rangle$ with $m=\pm 1,0$, all defined on the bond $b$. In a Fock space with the vacuum denoted as $
|\varnothing\rangle_b$, the singlet and triplet operators are defined as:
\begin{subequations}
\be
|s_b\rangle=\hats_b^\dagger|\varnothing\rangle_b
\ee 
\be
|t^m_b\rangle=\hatt^{m\dagger}_{b}|\varnothing\rangle_b,
\ee
\end{subequations}
with $\hats_b$ and $\hatt^m_{b}$ being bosonic operators. A boson number constraint 
\be\label{eq-cons}
\hat{s}_b^\dagger \hat{s}_b+\sum_{m=-1,0,1}\hatt^{m\dagger}_b\hatt^m_{b}=1
\ee
must also be satisfied on every bond. In terms of the singlet and the triplet operators defined above $\mathcal{H}_{\begin{gathered}\includegraphics[scale=0.14]{./t1.pdf}\end{gathered}}$ reads as,
\be 
\mathcal{H}_{\begin{gathered}\includegraphics[scale=0.14]{./t1.pdf}\end{gathered}}=-\frac{3}{4}J_h\sum_{b}\hats_b^\dagger \hats_b+\frac{1}{4}J_h\sum_{b}\sum_{\alpha=x,y,z} \hatt^{\alpha\dagger}_b\hatt^\alpha_{b}
\ee
where 
\bes
\be
\hatt^{x\dagger}=\frac{1}{\sqrt{2}}\left(\hat{t}^{-1\dagger}_b-\hat{t}^{1\dagger}_b\right)
\ee 
\be
\hatt^{y\dagger}=\frac{i}{\sqrt{2}}\left(\hat{t}^{-1\dagger}_b+\hat{t}^{1\dagger}_b\right)
\ee
\be
\hatt^{z\dagger}=\hat{t}^{0\dagger}_b.
\ee
\ees
Continuing, the next step involves rewriting the full Hamiltonian and expressing it in terms of the "coordinate" operator
\be 
\hat{Q}_{b}^{\alpha}=\frac{1}{\sqrt{2}}\left(\hat{t}_{i}^{\alpha\dagger}+\hat{t}_{i}^\alpha\right)
\ee
and its conjugate momentum operator
\be 
\hat{P}_{b}^{\alpha}=\frac{i}{\sqrt{2}}\left(\hat{t}_{i}^{\alpha\dagger}-\hat{t}_{i}^\alpha\right)
\ee
with $\alpha=x,y,z$.
Thus, the final form of the Hamiltonian, $\mathcal{H}$ containing $N_{\rm uc}$ unit-cells reads as,
\be\label{eq-MF}
\mathcal{H}\approx \mathcal{H}_{\text{MF}}= e_0 N_{\rm uc}+\frac{1}{2}\sum_{\bf{k}}\sum_{\alpha}\left[\lambda \hat{\bf{P}}_{\bf{k}}^{\alpha\dagger}\hat{\bf{P}}_{\bf{k}}^{\alpha}+\hat{\bf{Q}}_{\bf{k}}^{\alpha\dagger}\mathcal{V}_{\bf{k}}^\alpha\hat{\bf{Q}}_{\bf{k}}^{\alpha}\right].
\ee
In this context, $e_0$ is given by $e_0=-3J_h\overline{s}^2+\frac{3}{4}J_h+3\lambda\overline{s}^2-\frac{15}{2}\lambda$. Here, $\overline{s}$ represents the mean singlet amplitude on all the $J_h$ bonds, and $\lambda$ functions as the Lagrange multiplier employed to maintain the average satisfaction of the boson number constraint in \eqref{eq-cons}.
\bea
\hat{\bf{P}}_{\bf{k}}^{\alpha\dagger}=\left[\hat{P}_{1\bf{k}}^{\alpha\dagger}\text{ }\hat{P}_{2\bf{k}}^{\alpha\dagger}\text{ }\hat{P}_{3\bf{k}}^{\alpha\dagger}\right]\\
\hat{\bf{Q}}_{\bf{k}}^{\alpha\dagger}=\left[\hat{Q}_{1\bf{k}}^{\alpha\dagger}\text{ }\hat{Q}_{2\bf{k}}^{\alpha\dagger}\text{ }\hat{Q}_{3\bf{k}}^{\alpha\dagger}\right]
\eea
and 
\be
\mathcal{V}_{\bf{k}}^\alpha=\begin{bmatrix}
\lambda & \eta_{12} & \eta_{31}^\ast \\
\eta_{12}^\ast & \lambda & \eta_{23} \\
\eta_{31} & \eta_{23}^\ast & \lambda 
\end{bmatrix}  
\ee 
with
\bes
\be 
\eta_{12}=\frac{\overline{s}^2}{2}\left[-J_h-J_de^{-i\k\cdot\bf{a}_2}+J_t \left(e^{-i\k\cdot\bf{a}_1}+e^{-i\k\cdot\bf{a}_2}\right)\right]
\ee 
\be 
\eta_{23}=\frac{\overline{s}^2}{2}\left[-J_h-J_de^{i\k\cdot\bf{a}_1}+J_t \left(e^{i\k\cdot\bf{a}_1}+e^{i\k\cdot\left[\bf{a}_1-\bf{a}_2\right]}\right)\right]
\ee 
\be 
\eta_{31}=\frac{\overline{s}^2}{2}\left[-J_h-J_de^{-i\k\cdot\left[\bf{a}_1-\bf{a}_2\right]}+J_t \left(e^{i\k\cdot\bf{a}_2}+e^{-i\k\cdot\left[\bf{a}_1-\bf{a}_2\right]}\right)\right].
\ee 
\ees 
(the lattice vectors $\mathbf{a}_1=\sqrt{7}/2(\hat{x}+\sqrt{3}\hat{y})$ and $\mathbf{a}_2=\sqrt{7}\hat{x}$). Moreover, $\hat{P}^{\alpha\dagger}_{b\bf{k}}$'s and $\hat{Q}^{\alpha\dagger}_{b\bf{k}}$'s are the Fourier components of $\hat{P}^{\alpha\dagger}_{b}(\bf{r})$'s and $\hat{Q}^{\alpha\dagger}_{b}(\bf{r})$'s, respectively, i.e. $\hat{P}^{\alpha\dagger}_{b\bf{k}}=1/\sqrt{N_{\rm uc}}\sum_{\bf{k}}e^{i\mathbf{k}\cdot{\mathbf{r}}}\hat{P}^{\alpha\dagger}_{b}(\bf{r})$ and $\hat{Q}^{\alpha\dagger}_{b\bf{k}}=1/\sqrt{N_{\rm uc}}\sum_{\bf{k}}e^{i\mathbf{k}\cdot{\mathbf{r}}}\hat{Q}^{\alpha\dagger}_{b}(\bf{r})$.

The resulting $\mathcal{H}_{\text{MF}}$ is a system of three coupled harmonic oscillators, which is diagonalized to obtain 
\be
\mathcal{H}_{\text{MF}}=e_0 N_{\rm uc}+\sum_{\nu=1}^3\sum_{\bf{k}}\sum_{\alpha}\omega_{\nu}^{\alpha}(\k)\left(\gamma_{\k,\nu}^{\alpha\dagger}\gamma_{\nu}^{\alpha}(\k)+\frac{1}{2}\right)
\ee 
where $\gamma_{\k,\nu}^{\alpha}$ are renormalized triplon operators, and 
\be\label{eq-disp}
\omega_{\nu}^{\alpha}(\k)=\sqrt{\lambda\left(\lambda-\overline{s}^2\xi_{\nu}^{\alpha}(\k)\right)}
\ee
with
\be 
\xi_{\nu}^{\alpha}(\k)=\sqrt{-\frac{p_\mathbf{k}}{3}}\cos\left[\frac{1}{3}\cos^{-1}\left(\frac{3q_\mathbf{k}}{2p_\mathbf{k}}\sqrt{-\frac{3}{p_\mathbf{k}}}-\frac{2\pi}{3}m\right)\right]
\ee 
and
\bea
p_\mathbf{k}&=&-\left(|\eta_{12}|^2+|\eta_{23}|^2+|\eta_{31}|^2\right)\\
q_\mathbf{k}&=&2\text{Re}\left(\eta_{12}\eta_{23}\eta_{31}\right).
\eea

\begin{figure*}
\includegraphics[width=0.95\textwidth]{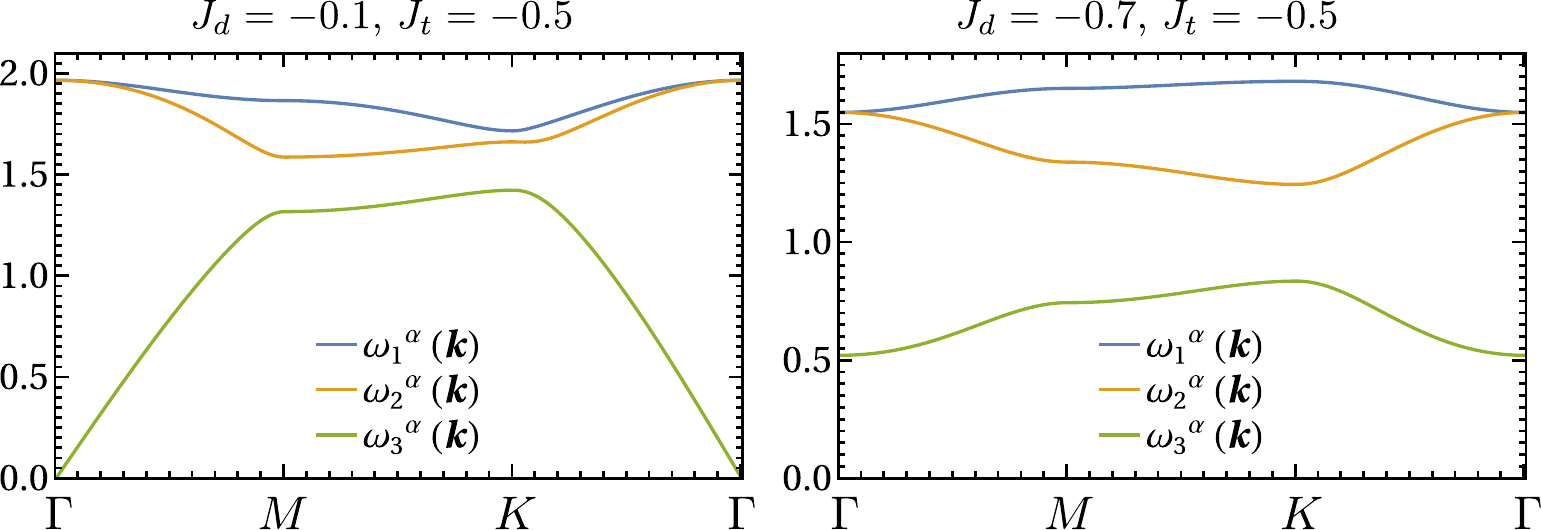}
\caption{he triplon dispersions [as in Eq.~\ref{eq-disp}] MLM in the gapless (left) and the gapped (right) phase.} \label{fig-disp}
\end{figure*}

The system's ground state is determined by the vacuum of the quasi-particles, denoted as $\gamma_{\k,\nu}^{\alpha}$. Consequently, the ground state energy per site of the system is expressed as follows,
\be\label{eq-gs}
e_g=\frac{e_0}{6}+\frac{1}{12N_{\rm uc}}\sum_\nu\sum_{\bf{k}}\sum_{\alpha}\omega_{\nu}^{\alpha}(\k)\ee 
The unknown mean-field parameters, $\lambda$ and $\overline{s}^2$ are determined by minimizing $e_g$, which leads to the following self-consistent equations
\bes\label{eq-sc1}
\be
\lambda=J_h+\frac{1}{12N_{\rm uc}}\sum_\nu\sum_{\bf{k}}\sum_{\alpha}\frac{\lambda \xi_{\nu}^{\alpha}(\k)}{\omega_{\nu}^{\alpha}(\k)}
\ee
\be 
\overline{s}^2=\frac{5}{2}-\frac{1}{12N_{\rm uc}}\sum_\nu\sum_{\bf{k}}\sum_{\alpha}\frac{2\lambda-\overline{s}^2 \xi_{\nu}^{\alpha}(\k)}{\omega_{\nu}^{\alpha}(\k)}.
\ee
\ees

This formulation provides two distinct physical solutions depending on whether the triplon dispersions exhibit a gap or are gapless. The $\mathcal{H}_{\text{MF}}$ encompasses nine triplon dispersions. In the scenario where the minimum of the lowest dispersion in the Brillouin zone is strictly greater than zero (as depicted in Fig.\ref{fig-disp}), it indicates the presence of an energy gap protecting the dimerized ground state against triplon excitations. In this 'gapped' dimerized phase, Eqs.\ref{eq-sc1} is applicable in the given form.

However, as $J_d$ weakens, the triplon gap may close at a specific point $\mathbf{q}$ in the Brillouin zone. In other words, $\omega_{\nu}^{\alpha}(\mathbf{q})=0$ for certain lower triplon branches. If this occurs, the corresponding terms in Eqs.~\ref{eq-sc1} for $\mathbf{k} = \mathbf{q}$ become singular, leading to triplon condensation described by the condensate density, $n_c$, a third unknown in the problem. Now, a third equation is introduced, representing the condition of gaplessness. Based on our calculations, we observe $\omega_{\nu'}^{\alpha}(\mathbf{q})=0$ at $\mathbf{q} = (0, 0)$ for $\alpha = x, y, z$, and $\nu'= 3$ (Fig.~\ref{fig-disp}). The revised equations applicable to the gapless case are given by
\bes  \label{eq-sc2}
\be
\lambda=\overline{s}^2\xi_{\nu'}^{\alpha}(\mathbf{Q})
\ee
\be 
\overline{s}^2=\frac{5}{2}-n_c-\frac{1}{12N_{\rm uc}}\sum_{\nu\ne\nu'}\sum_{\bf{k}}\sum_{\alpha}\frac{2\lambda-\overline{s}^2 \xi_{\nu}^{\alpha}(\k)}{\omega_{\nu}^{\alpha}(\k)}-\frac{1}{12N_{\rm uc}}\sum_{\bf{k}\ne\mathbf{Q}}\sum_{\alpha}\frac{2\lambda-\overline{s}^2 \xi_{\nu'}^{\alpha}(\k)}{\omega_{\nu'}^{\alpha}(\k)}.
\ee
\be 
n_c=\left(1-\frac{J_h}{\lambda}\right)\overline{s}^2-\frac{1}{12N_{\rm uc}}\sum_{\nu\ne\nu'}\sum_{\bf{k}}\sum_{\alpha}\frac{\overline{s}^2 \xi_{\nu}^{\alpha}(\k)}{\omega_{\nu}^{\alpha}(\k)}-\frac{1}{12N_{\rm uc}}\sum_{\bf{k}\ne\mathbf{Q}}\sum_{\alpha}\frac{\overline{s}^2 \xi_{\nu'}^{\alpha}(\k)}{\omega_{\nu'}^{\alpha}(\k)}.
\ee
\ees

\begin{figure*}
\includegraphics[width=0.95\textwidth]{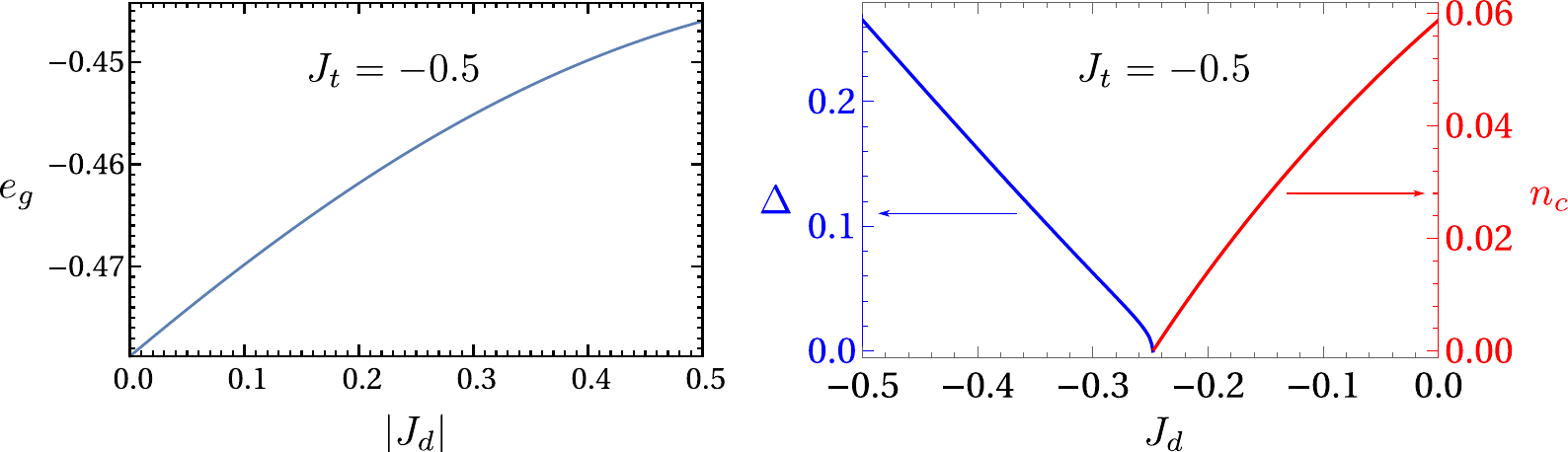}
\caption{The ground state energy, $e_g$, singlet triplet gap, $\Delta$, and condensation order parameter, $n_c$, as a function of $J_d$.} \label{fig-gap}
\end{figure*}

In Fig.~\ref{fig-gap}, we dispaly the ground state energy per spin obtained from Eq.~\ref{eq-gs} and the spin gap, $\Delta$, which is the minimum of $\omega_{\nu}^{\alpha}(\k)$, and $n_c$ calculated using Eqs.~\ref{eq-sc1} and~\ref{eq-sc2}. 

\section{Further results from DMRG}
\begin{figure}
\includegraphics[width=0.8\columnwidth]{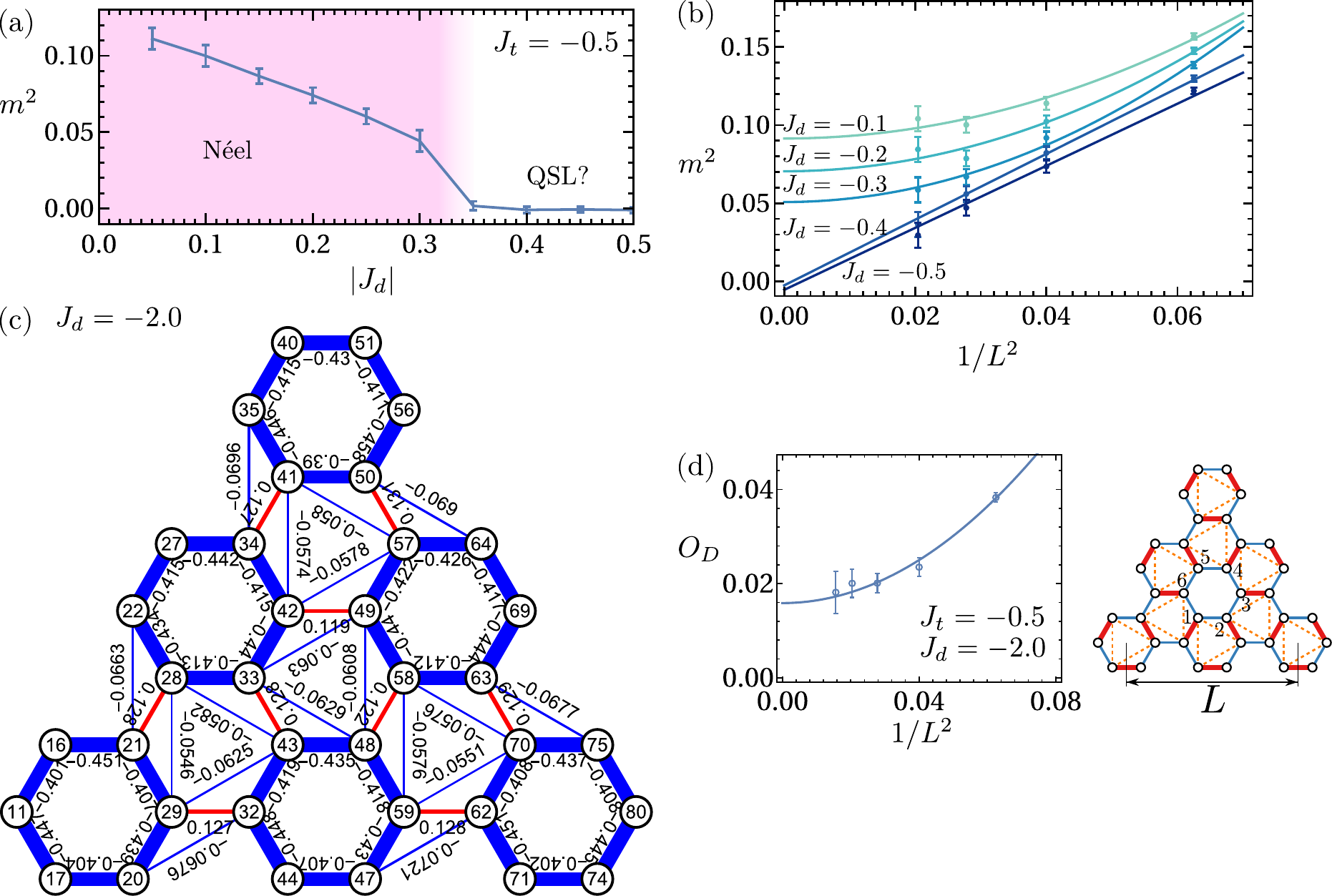}
\caption{(a) Finite size scaled N\'eel order parameter ~\eqref{eq-m2} as a function of $J_d$ for $J_t=-0.5$ and $J_h=1$. (b) Finite size scaling of the N\'eel order parameter for $(J_h,J_t)=(1,-0.5)$ for different values of $J_d$. (c) We depict the spin correlations across the nearest-neighbor bonds for $(J_h,J_t,J_d)=(1,-0.5,-2.0)$ with the thickness of the lines being proportional
to the respective correlation. Blue lines indicate negative (antiferromagnetic) correlations while orange lines imply positive (ferromagnetic) correlations. The edge sites are removed. (d) Finite-size scaling of the dimerization order parameter ~\eqref{eq-OD} for $J_d=-2$.} \label{fig-DMRG}
\end{figure}

We further investigated the system using density matrix renormalization group (DMRG) methods~\cite{White1992}. We do our calculations only for $J_t=-0.5$. The DMRG calculations are performed utilizing the iTENSOR library~\cite{itensor}. We study various cluster sizes, reaching up to $216$ sites, and maintaining a bond dimension up to $\sim 4000$ during the renormalization process. Approximately 50 sweeps were performed until the ground-state energy converged within an error of $\sim 10^{-5}$. The results are summarized in Fig.~\ref{fig-DMRG}. The cluster geometry is depicted in the right panel of Fig.~\ref{fig-DMRG} (d), featuring open edges that favor dimerization of the $J_h$ hexagons. 

To identify the N\'eel order, we calculate the order parameter as
\be\label{eq-m2}
m^2(L)=\frac{1}{6^2N_{\text{uc}}^2}\sum_{\mathbf{R},\mathbf{R}^\prime}\sum_{i,j=1}^6(-1)^{i+j}\langle\S_i(\mathbf{R})\cdot\S_j(\mathbf{R}')\rangle,
\ee
where $i$ and $j$ run over the basis indices [convension indicated in Fig.~\ref{fig-DMRG} (d)], $\mathbf{R}$ and $\mathbf{R}'$ denote the position of the unit cells, and $N_\text{uc}$ is the total number of unit cells. This quantity exhibits good finite-size scaling with terms proportional to $1/L^2$ and $1/L^4$ Fig.~\ref{fig-DMRG} (b)], where $L$ is the linear dimension of the system. Fig.~\ref{fig-DMRG} (a) shows that the scaled N\'eel order parameter, $m^2$, gradually decreases as $|J_d|$ increases. At $|J_d|=0.35$, $m^2\approx 0$, indicating the appearance of a magnetically disordered phase.

The dimerization order parameter, given by
\be\label{eq-OD}
O_D(N_\text{uc})=\frac{1}{3N_{\text{uc}}}\sum_{\mathbf{R}}\left|\sum_{\langle ij\rangle}^{{\begin{gathered}\includegraphics[scale=0.14]{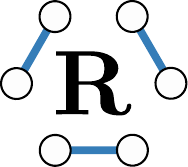}\end{gathered}}}\langle\S_i\cdot\S_j\rangle-\sum_{\langle ij\rangle}^{{\begin{gathered}\includegraphics[scale=0.14]{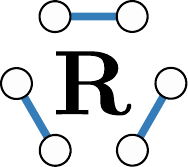}\end{gathered}}}\langle\S_i\cdot\S_j\rangle\right|,
\ee  
was not consistently obtained for small $|J_d|>0.35$, making it challenging to reliably detect the onset of the dimerized phase. However, for larger values of $|J_d|$, e.g., $J_d=-2.0$, $O_D$ shows a clear dimerization tendency [see Fig.~\ref{fig-DMRG} (d)]. The degree of dimerization, however, is not markedly pronounced, akin to the trimerization observed in spin-$1$ systems~\cite{Changlani-Spin-1_Kagome}. In Fig.~\ref{fig-DMRG} (c), we portray the the NN bond energy $\langle\mathbf{S}_i\cdot\mathbf{S}_j\rangle$ for $(J_h,J_t,J_d)=(1,-0.5,-2.0)$ on a cluster with $L=4$. The red and blue bonds denote the positive and negative bond energies. Note from the values of the bond energies, the $J_h$ hexagons are clearly dimerized. 


%